\documentclass[12pt]{iopart}
\usepackage{iopams}
\usepackage{graphicx}
\usepackage{xcolor}

\usepackage{algorithm,algpseudocode}
\algnewcommand{\Inputs}[1]{%
	\State \textbf{Inputs:}
	\Statex \hspace*{\algorithmicindent}\parbox[t]{.8\linewidth}{\raggedright #1}
}
\algnewcommand{\Initialize}[1]{%
	\State \textbf{Initialize:}
	\Statex \hspace*{\algorithmicindent}\parbox[t]{.8\linewidth}{\raggedright #1}
}
\algnewcommand{\Algorithm}[1]{%
	\State \textbf{Algorithm:}
	\Statex \hspace*{\algorithmicindent}\parbox[t]{.8\linewidth}{\raggedright #1}
}

\def\doubleunderline#1{\underline{\underline{#1}}}

\begin{document}

	\title[Description of collective magnetization processes with machine learning models]{Description of collective magnetization processes with machine learning models}
	\author{Alexander Kornell$^{1,2,*}$, Lukas Exl$^{3,4}$, Leoni Breth$^{1,2}$, Johann Fischbacher$^{1,2}$, Alexander Kovacs$^{1,2}$, Harald Oezelt$^{1,2}$, Markus Gusenbauer$^{1,2}$, Masao Yano$^{5}$, Noritsugu Sakuma$^{5}$, Akihito Kinoshita$^{5}$, Tetsuya Shoji$^{5}$, Akira Kato$^{5}$, Norbert J. Mauser$^{3}$, Thomas Schrefl$^{1,2}$}
	
	\address{$^1$ Christian Doppler Laboratory for Magnet design through physics informed machine learning, Viktor Kaplan-Straße
		2E, 2700 Wiener Neustadt, Austria}
	\ead{alex.kornell@gmx.at}
	\address{$^2$ Department for Integrated Sensor Systems, Danube University Krems, Viktor Kaplan-Straße 2E, 2700 Wiener
		Neustadt, Austria}
	\address{$^3$ Research Platform MMM Mathematics-Magnetism-Materials c/o Fak. Mathematik, University of Vienna, Oskar-Morgenstern-Platz 1, 1090 Vienna, Austria}
	\address{$^4$ Wolfgang Pauli Institute c/o Fak. Mathematik, University of Vienna, Oskar-Morgenstern-Platz 1, 1090 Vienna, Austria}
	\address{$^5$ Advanced Materials Engineering Div., Toyota Motor Corporation, 1200, Mishuku Susono, Shizuoka 410-1193 Japan}

	\begin{abstract}
	This work introduces a latent space method to calculate the demagnetization reversal process of multigrain permanent magnets. The algorithm consists of two deep learning models based on neural networks. The embedded Stoner-Wohlfarth method is used as a reduced order model for computing samples to train and test the machine learning approach. The work is a proof of concept of the method since the used microstructures are simple and the only varying parameters are the magnetic anisotropy axes of the hard magnetic grains.
	A convolutional autoencoder is used for nonlinear dimensionality reduction, in order to reduce the required amount of training samples for predicting the hysteresis loops. We enriched the loss function with physical information about the underlying problem which increases the accuracy of the machine learning approach. A deep learning regressor is operating in the latent space of the autoencoder. The predictor takes a series of previously encoded magnetization states and outputs the next magnetization state along the hysteresis curve. To predict the complete demagnetization loop, we apply a recursive learning scheme.

	\end{abstract}
	\submitto{Machine Learning: Science and Technology}
	 
	\maketitle
	
	\section{Introduction}
	
	High performance permanent magnets play a key role in many technologies of today's life. They are necessary for cell phones, consumer electronic devices, hybrid and electrical vehicles and sustainable energy production. Especially the last two mentioned markets are emerging quickly and therefore the need for permanent magnets will grow significantly in the future \cite{Meszaros2020, Ntanos2018, Yang2017}. 
	The state of the art intermetallic phase of high performance permanent magnets is Nd$_2$Fe$_{14}$B. Neodymium (Nd) is considered a critical element because of high demand and possible supply risks \cite{crm2020}. Therefore, it is essential to avoid the use of Nd while keeping the required magnetic properties for the target applications \cite{Binnemans2018}. 
	
	High performance magnets are usually clustered by the magnetic field they are able to provide and their resistance against opposite magnetic fields. These macroscopic properties are reflected in the coercive field, the remanence and the energy density product and emerge from microscopic magnetic properties and the microstructure of the magnet \cite{Fischbacher2018}. Looking at this dependence it is clear that the development of novel high performance magnets must be based on a thorough understanding of the microscopic properties. The involved range of length scales provides a particular challenge towards exact estimations at the device level. Physical experiments are often costly and sometimes infeasible and that is why accurate simulation strategies are necessary. This work contributes to the multiscale description of permanent magnets, introducing a methodology to predict magnetization reversal at the level of the granular microstructure of a magnet.   
	
	The continuum theory of micromagnetism operates on a length scale, which is small enough to model the microscopic properties of interest. Nevertheless, it is worth mentioning that classical numerical algorithms based on this theory are either costly in terms of computation time or require crude approximations to bridge the length scales. These are the reasons why modern research on this topic is now exploring algorithms based on machine learning as an efficient alternative \cite{schaffer2021machine, exl2021prediction, Kovacs2019, Exl2018}. Accurate machine learning models, however, can only be developed if there is enough training data from experiments or simulations. Within this approach the numerical algorithms, which are able to solve the fundamental differential equations of micromagnetism, are the basis of the models.
	
	Successfully applying machine learning to calculate the figures of merit for high performance magnets enables to pass microscopic properties through the scales. This leads to a new approach of material design, which will provide an efficient way to explore novel material combinations for rare-earth-free permanent magnets.	
	
    \section{Related Works}
	
	Our proposed approach is a latent space method. This means that one uses a dimensionality reduction to create embedded images of the original input. Since our method is data-driven, we use a convolutional autoencoder to span the lower dimensional latent space. This is a deep learning model. Moreover, the regressor is operating on the embedded images since they are of lower dimension. 
	The initial idea of this approach is coming from \cite{Kim2018}. Kim et al. present a generative model for synthesize fluid simulations based on a convolutional autoencoder and an integrator operating in the latent space. Reconstructing the velocity fields can be done approximately 700x faster than re-simulating the data with the numerical algorithm. 
	
	In the field of micromagnetism, the work \cite{Exl2018} proposes a machine learning method to identify the nanocrytalline grain with the smallest switching field. However, in this work the latent space is constructed by predefined attributes of all grains composing the microstructure. In dynamic micromagnetism, Kovacs et al. \cite{Kovacs2019} recently proven that a latent space method can predict the magnetic response of thin film magnets. They successfully use a machine learning regressor within the latent space of convolutional autoencoder for the time integration of the Landau-Lifshitz-Gilbert (LLG) equation \cite{Kovacs2019}. Finally, Schaffer et al. are comparing different dimensionality reduction methods for constructing the latent space for a reduced order regressor \cite{schaffer2021machine}.        
	
	\section{Methods}
	
	\subsection{Embedded Stoner-Wohlfarth Method}
	
	Our proposed method relies on deep learning. The main problem of these types of machine learning models is that they heavily rely on a large dataset. In that sense, it is crucial to use a very efficient simulation technique to be able to calculate a rich enough dataset. This is the reason for choosing a reduced order model for the necessary micromagnetic simulations to train and test the machine learning method. The benefit of the \textit{embedded Stoner-Wohlfarth method} \cite{Mehofer2017} is that it can capture low level information of interacting grains while keeping the computation effort on the minimum. 
	
	The standard \textit{Stoner-Wohlfarth model} describes a magnetic particle which is small enough to only have a single domain and it has an elliptical shape. Although the particle must be very small, it has to be large enough to have a net magnetization. 
	In this simple setup, important properties like the switching field $H_n$ can be calculated analytically. In detail, the external field required to switch the magnetization of the Stoner-Wohlfarth particle is given by \cite{DissOezelt2018}: 
	
	\begin{equation}\label{eq:StonerWohlfarthtSwitchingField90}
	H^{0}_{n} = \frac{2(K_u + K_d)}{\mu_0 M_s} \textrm{,}
	\end{equation}

	assuming that the external field is exactly anti-parallel to the easy axis $\mathbf{k}$ of the magnet, where $\mathbf{k}$ is a unit vector along the magnetocrystalline anisotropy axis. Here $K_u$ is the anisotropy constant and $K_d = \frac{1}{2} \mu_0 M_s^2 (N_{\bot} - N_{\parallel})$ is the effective anisotropy constant coming from the demagnetization field \cite{Kronmueller1987a} with $N_{\parallel}$ and $N_{\bot}$ denoting the demagnetization factors of the magnetization parallel and perpendicular to the easy axis. This field is also referred to as the nucleation field of the magnet which is formally defined as the minimal field which causes the magnetization to rotate out of its preferred axis.
	
	However, if the angle between the easy axis and the applied field is $0 < \Theta_h \leq \pi/2$, the field to switch the magnetization of the grain is given by \cite{Exl2018}: 
	
	\begin{equation}\label{eq:StonerWohlfarthtSwitchingField}
	H_{n} = H_{n}^{0} \big(\cos^{2/3}(\Theta_h) + \sin^{2/3}(\Theta_h)\big)^{-3/2} \textrm{.}
	\end{equation}

	\begin{figure} 
		\centering
		\includegraphics[scale=0.36]{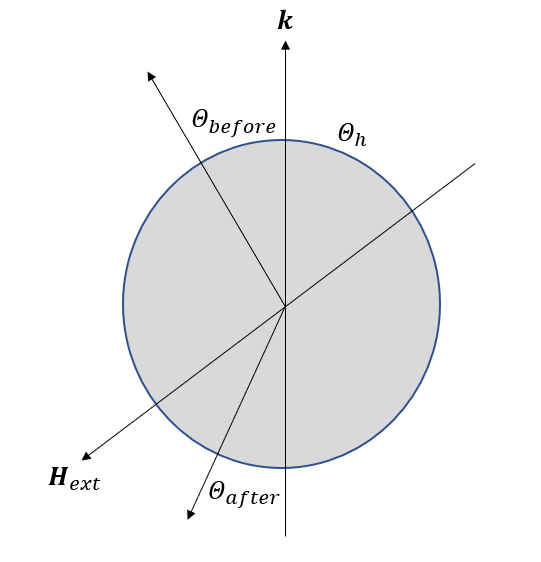}
		\caption{\label{fig:stonerwohlfarthparticle} Sketch of Stoner-Wohlfarth particle.}
	\end{figure}  

	In addition, there is a reversible rotation of the magnetization of the grain, whenever the applied field is not exactly anti-parallel to the easy axis of the grain. Before the grain reverses its magnetization irreversible, the magnetization rotates away from the easy axis until a critical angle is reached. At this point the magnetization spontaneously changes its direction towards a near direction of the external field. This process is schematically shown in Figure \ref{fig:stonerwohlfarthparticle}.  Finally, $\Theta$ converges to $\Theta_h$ if $\mu_0M_sH_{ext} >> 2(K_u+K_d)$. The exact solution can be found in the work \cite{Kalezhi2011} of J. Kalezhi and it will be used for the numerical calculations of this work since it is very important in the practical use of the \textit{embedded Stoner-Wohlfarth method}. 
	
	The \textit{embedded Stoner-Wohlfarth method} distributes virtual Stoner-Wohlfarth particles in each grain whereby each particle has the same intrinsic magnetic properties as the grain in which it is embedded. Within a grain, all Stoner-Wohlfarth particles have the same easy axis. Therefore the intrinsic switching field according to equation (\ref{eq:StonerWohlfarthtSwitchingField}) is the same for all particles within the grain. However, the field acting on the particle varies since it is a sum of the external field and the local demagnetization fields of the other grains. In addition, the switching field of a grain is defined by the reversal of the embedded Stoner-Wohlfarth particles: It is expected that the complete grain reverses its magnetization if the total field at one of the particles is greater than the intrinsic switching field. This is justified because grains of a Nd$_2$Fe$_{14}$B magnets contain no sites for domain wall pinning. In an external field, an initially reversed domain will grow and the entire grain will become reversed.
	
	To avoid any naming complications, the total field $\mathbf{H}_{tot}$ is introduced as the magnetic field at a Stoner-Wohlfarth particle coming from the externally applied field and the demagnetization field of the grains. This total magnetic field can be calculated by \cite{Exl2018}:
	\begin{equation}\label{eq:ROMHtot}
	\mathbf{H}_{tot}(\mathbf{x}) = \mathbf{H}_{ext} + \mathbf{H}_{demag}(\mathbf{x}) \textrm{.}
	\end{equation}    
	This equation explicitly uses the spatial coordinate $\mathbf{x}$ to underlie that the total field $\mathbf{H}_{tot}$ is eventually different at each discretization point of the system although the external field $\mathbf{H}_{ext}$ is constant within the volume of interest. Here we neglect ferromagnetic exchange interactions between the grains assuming that the grains are separated by a non-magnetic intergranular phase. 

	\begin{algorithm}[ht]
	\caption{Reduced Order Model}
	\label{alg:reducedordermodel}
	\begin{algorithmic}[1]
		\Inputs{$\mathbf{M}_0$, $M_s$, $K_a$, $K_d$, $\mathbf{k}$, $\mathbf{h}_{ext}$ \newline}
		\Initialize{MAXITER, $\Delta H_{ext}$, $H^{init}_{ext}$, $H_{ext}^{max} > 0$  \\ $H_{ext} \gets H^{init}_{ext}$ \\ $\mathbf{M}[H^{init}_{ext}][0] \gets \mathbf{M}_0$ \newline}
		
		\While{$H_{ext} \leq H_{ext}^{max}$}
		\For{i = 1 to MAXITER}
		\For {SW-particle j in microstructure}
		\State {$\mathbf{H}_{demag} \gets$ calculate demagnetization field at SW-particle j}
		\State {$\mathbf{H}_{tot} \gets \mathbf{H}_{demag} + H_{ext} \mathbf{h}_{ext}$}
		
		\State {$\Theta_{h_{\textrm{tot}}} \gets$ angle between total field and $\mathbf{k}$}
		
		\State{$H_{switch} \gets$ equation (\ref{eq:StonerWohlfarthtSwitchingField})}

		\If{$||\mathbf{H}_{tot}|| > H_{switch}$}
		\State {switch all SW-particles of grain}
		\State {$\mathbf{M}[H_{ext}][j] \gets$ \cite{Kalezhi2011} adapted to magnetization}
		\Else
		\State {$\mathbf{M}[H_{ext}][j] \gets$ \cite{Kalezhi2011} adapted to magnetization}
		\EndIf
		\EndFor
		\If{no particle switched}
		\State {break}
		\EndIf
		
		\EndFor
		\State {$H_{ext} \gets H_{ext} + \Delta H_{ext}$}
		\EndWhile
		
		\State {\textbf{return} $\mathbf{M}$}
		
	\end{algorithmic}
    \end{algorithm}

	Algorithm \ref{alg:reducedordermodel} describes the \textit{embedded Stoner-Wohlfarth method} in pseudo code. To clarify uncertainties, it is worth to mention that the strength of the external field is treated as a scalar value $H_{ext}$ while its direction is given by the unit vector $\mathbf{h}_{ext}$ which is pointing in the opposite direction of the initial saturation field. The index $j$ selects the magnetization or the total field of the $j$-th particle. \newline 
	The first step to simulate the magnetization reversal processes is to create a multigrain structure where each grain has slightly different intrinsic magnetic properties since this is the case in nature too. In this setup, the difference is restricted to the easy axes $\mathbf{k}$ because the model is strictly built of cubic grains. However, these properties of the model are the inputs of Algorithm \ref{alg:reducedordermodel} in addition to an initial state of the magnetization. \newline
	Next, the size of the external field steps is defined. Furthermore, an iteration for each field step is started where the total field is calculated at the beginning. In addition, the magnetization is computed by the equations of the \textit{Stoner-Wohlfarth model} where a check if the total field exceeds the switching field is needed. If this is the case, it is necessary to continue the iteration since the total field for the other grains changes significantly. Lastly the algorithm can move on to the next external field step if the system reaches a stable state where no more switching fields are exceeded. 
	
	\subsection{Demagnetization Field}
	
	A very important computational step of Algorithm \ref{alg:reducedordermodel} is the calculation of the demagnetization field in line 6. Within the framework of the \textit{embedded Stoner-Wohlfarth method}, it is assumed that each grain is uniformly magnetized and that they are separated by an isolating boundary phase. Because $\nabla' \cdot \mathbf{M}(\mathbf{x}') = 0$ within each grain, there are no field sources inside a grain. As a consequence, the demagnetizing field can be expressed as sum over surface integrals 
	\cite{Exl2018}: 
	
	\begin{equation}\label{eq:HdemagROM}
	\mathbf{H}^{eSW}_{demag}(\mathbf{x}) = \frac{1}{4\pi} \sum_i^{N_g} \sum_j^{N_{s_i}} \int_{\partial V'_{ij}} \mathbf{M}\cdot \mathbf{n}_{ij} \frac{(\mathbf{x}-\mathbf{x}')}{||\mathbf{x}-\mathbf{x}'||^3} dS'_{ij} \textrm{.}
	\end{equation}
	
	where $N_g$ is the total number of grains and $N_{s_i}$ is the number of surfaces of the $i$-th grain. The vector $\mathbf{n}_{ij}$ is the outer normal to surface $k$ of grain $j$.
	
	Finally, it is possible to replace the surface integral by integrals over all edges of the microstructure by again using the fact that the grains are uniformly magnetized. Guptasarma and Singh \cite{Guptasarma1998} reported the necessary formulas for calculating the demagnetization field for uniformly magnetized polyhedrons. 
	It is worth to mention that the necessary equations can be implemented efficiently in a vectorized formulation. This is especially important when the algorithm is implemented in a high level programming language like Python. Without any further details, the final results are mainly consisting of the following matrix vector product:  
	   
	\begin{equation}\label{eq:HdemagROMSumVector}
	\vec{H}^{ROM}_{demag} = \doubleunderline{B} \cdot \vec{M}  \textrm{,}
	\end{equation}
	
	where $\vec{M}$ and $\vec{H}^{ROM}_{demag}$ hold the magnetization and the demagnetizing field of all grains, respectively. The matrix $\doubleunderline{B} \in R^{3N_p \times 3N_g}$ contains the spatial components of the structure, where $N_p$ is the total number of Stoner-Wohlfarth particles. One of the main benefits of this vectorized equation (\ref{eq:HdemagROMSumVector}) of the demagnetization field is that it can be implemented very efficiently on graphic processors since it is possible to highly parallelize the computation \cite{Mehofer2017}.  
	
	Nonetheless even an optimized algorithm has the drawback that the calculation of the demagnetization field at each iteration has a theoretical runtime of the order $\mathcal{O}(N_g N_p)$ because it needs $N_{p}$ multiplications for every grain. Going back to physics, one notices that the demagnetization field decreases with increasing distance. This clearly indicates that neighboring grains have an higher impact to the demagnetization field than elements which are far away from each other. This means that grains which are far away from the point of interest can be clustered without a big lack in accuracy. 
	A similar approximation is successfully applied by the Barns-Hut tree method which is used to solve N-body problems in astrophysics or electrodynamics \cite{Iwasawa2020, Gan2013}. The tree code reduces the theoretical run time to the order $\mathcal{O}(nlog(n))$ and there are libraries available which implement this method. To use this idea, one needs to start by enumerating the grains by their distances and add their values accordingly to $\vec{M}$ and $\doubleunderline{B}$. This enables to calculate the matrix vector product by hierarchical matrices which are referred to as $\mathcal{H}^2$-matrices. The main benefit of the $\mathcal{H}^2$-matrices is that the calculation of the product can even be done with nearly a linear order \cite{Boerm2006}.  One example is the Python package h2tools \cite{Mikhalev2015}. 
	
	\subsection{Machine Learning Methods}
	
\begin{figure}[!ht]
		\centering
		\includegraphics[scale=0.39]{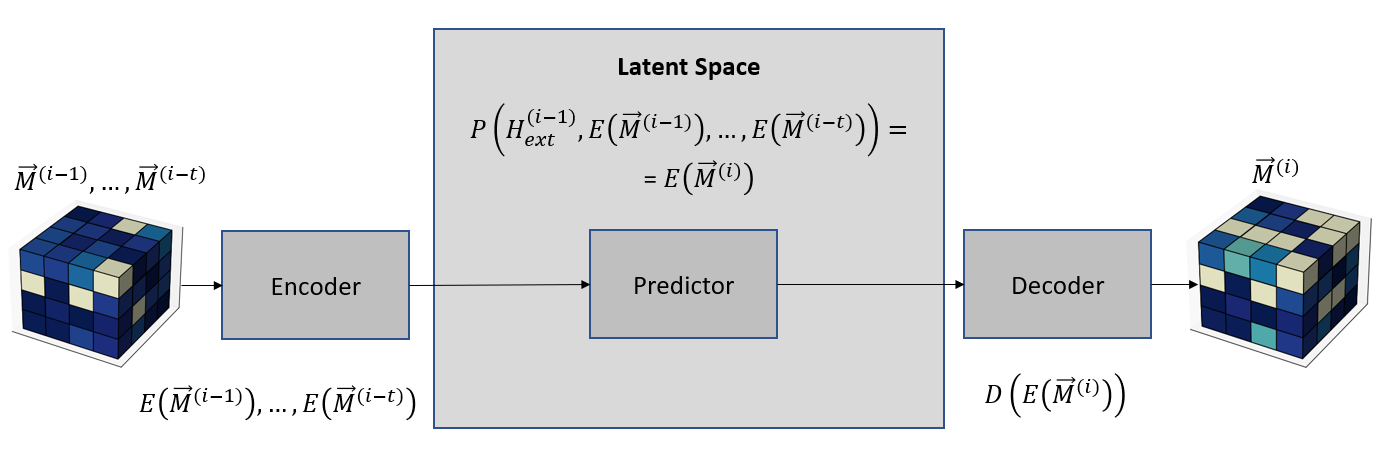}
		\caption{\label{fig:mlmodel} Schematic overview of the method to predict the next step of the hysteresis by using an autoencoder and a predictor in the latent space.}
	\end{figure}

	The machine learning approach proposed in this work consists of two models: 1) an convolutional autoencoder for dimensionality reduction, 2) a neural network regressor for predicting magnetization states. In addition, the predictor is applied to the latent space of the autoencoder. Figure \ref{fig:mlmodel} outlines the method visually. 
	
	The algorithm starts by using the encoder of the autoencoder to obtain embedded images of the magnetization at the $i-t, ..., i-1$ steps of the demagnetization curve. Furthermore, the predictor model takes these images in the latent space and the external field of the $(i-1)$-th step of the hysteresis as the input to predict the magnetization in the latent space at the $i$-th step of the demagnetization curve. Finally the decoder must be used to transform the new magnetization state back to the real space. This predictor can be used to estimate the complete hysteresis of a magnetic material by only using the first $k$ magnetization states since it can be applied to its own predictions. In total this represents a method where the reduced order model is replaced by a machine learning model and this should lead to a substantial speedup. However, the regressor must calculate accurate results that one can use this method. 
	
	\subsubsection{Empirical Risk Minimization}
	
	The machine learning models used in this work are trained by supervised learning. The goal of this technique is to establish models which can reliably predict a variable $y \in \mathcal{Y}$, given an input $x \in \Omega$. A commonly used technique to obtain such models is called \textit{emprical risk minimization}. Given the empirical risk $\mathcal{L}_z$ and the loss function $\mathit{L}(\hat{y}, y)$, the empirical risk minimization problem can be defined as
	\cite{Schaffer2021}: 
	
	\begin{equation}\label{eq:ERM}
	\hat{f}_{\mathit{H}, z} = {\arg\min}_{f \in \mathit{H}}\{ \mathcal{L}_z(f)\}  = {\arg\min}_{f \in \mathit{H}}\left\{ \frac{1}{m} \sum^m_{i=1} \mathit{L}(\hat{y}_i, y_i)\right\} \textrm{,}
	\end{equation}

	In detail, a loss function takes the predicted value $\hat{y}$ and the ground truth $y$ and calculates a difference which is important for the underlying problem. It is necessary to understand that there is no 'one fits all' loss function because the importance of the underlying error can vary with the actual problem. However, some commonly used examples are the mean squared loss, the cross entropy loss and the hinge loss. Furthermore, it is possible to enrich the loss function by information of the underlying problem \cite{Kovacs2019}. 
	One of the main topics of this work is to engineer appropriate loss functions to predict the magnetization reversal process. This is done by including parts of micromagnetism, like the preservation of the norm of the magnetization vector, to this function. Furthermore, recursive learning of the predictor is realized by the loss function.
	
	However, training a machine learning model with \textit{empirical risk minimization} is basically a large algebraic  minimization problem. Since it is expected that the dataset is large, one usually uses gradient based minimization techniques. 
	Although the standard stochastic gradient descent method (SGD) is already a viable algorithm for training a machine learning model, there are ways to improve the algorithm by using additional information of the function to minimize. One example is the Adam algorithm which uses the momentum of the gradient and an adaptive learning rate to update the property of interest. This method converges faster than the classical stochastic gradient descent method and it has also a higher probability to escape local minima if the loss surface is noisy \cite{Kingma2017}. However, very commonly used machine learning models for dimensionality reduction are convolutional autoencoders based on neural networks. For that type of models there is a further improvement to the Adam algorithm which is named the Nadam method. The Nadam method essentially speeds up the training of a convolutional autoencoder by reducing the number of epochs needed to converge to a minimum \cite{Dozat2016}. 
	
	\subsubsection{Convolutional Autoencoder}
	
	Formally an autoencoder is an unsupervised neural network which tries to copy the input to its output \cite{Schaffer2021}. However, the interesting part is that there is one central intermediate layer which has a different dimension than the input layer. This technique is commonly used for dimensionality reduction or feature extraction. The focus of this work lays on the first case where the intermediate layer compresses the data to the most important features which are necessary to obtain the original state, assuming that the method is successfully reducing the dimension of the underlying problem.
	
	An autoencoder consists of two parts, namely the encoder and the decoder. The encoder maps the input data to a latent space where the dimension of this space is defined by the central intermediate layer. In addition, the decoder is able to decode the latent space representation of the input back to the real space. Since the representation in the latent space is unknown upfront, both parts of the autoencoder are trained together where the loss is usually calculated in the real space and the weights and biases of both models are updated at once. The following equation describes the empirical risk of an autoencoder: 
	
	\begin{equation}\label{eq:riskautoencoder}
	\mathcal{L}_{AE} = \frac{1}{m}\sum_{k=1}^m \mathit{L}(D(E(x_k)), x_k)  \textit{,}
	\end{equation}

	where $E: \Omega \rightarrow \mathcal{F}_\Omega$ is the encoder and $D: \mathcal{F}_\Omega \rightarrow \Omega$ is the decoder. In addition, the space $\mathcal{F}_\Omega \subset R^d$ is the latent space. When using this technique for dimensionality reduction, the dimension of the latent space is lower than the dimension of the real space. In that case one speaks of an undercomplete autoencoder. On the other hand an overcomplete autoencoder would have a latent space of higher dimension than the original space.
	
	Convolutional autoencoders are neural networks which utilize convolutional layers instead of dense layers. Convolutional layers restrict the input for each neuron to just a few neurons of the last layer which are geometrically nearby each other. This is done by using a filter or kernel with a specific size for calculating the output of a layer. In detail, the filters given by matrices with a previously defined dimension are applied to all input channels of the layer and the depth of the output image is equal to the number of filters applied within the layer \cite{Buduma2017}. This essentially means that a convolutional layer transforms a volume of values into another volume of values which eventually has a different dimension. Finally, the activation function is applied component-wisely. As a consequence the total number of trainable parameters is drastically reduced. This technique allows to identify local characteristics of the image at the first layers while global characteristics are recognized within the deeper layers of the network. Although this method was initially developed for image recognition, it also shows its strength when applied to physical problems because local interactions eventually have an other impact to the system than global ones. 
	
	In addition, convolutional layers can include downsampling directly within the layer by using a stride. This value defines how far the filter is moved from one calculation to the next. For example a stride of two means that the filter is shifted by two units. This procedure reduces the dimensionality already within the convolutional layer because the number of elements in the output matrix is roughly the number of total input values divided by the stride in each dimension and for each filter. The main benefit of this procedure is that it introduces trainable parameters for the dimensionality reduction step. This is not the case for pooling layers since they are consisting of fixed operations. Nevertheless, there is a drawback because the increased number of weights and biases require more computational power to be trained. However, recent researches show that there are applications where convolutional layers with strides outperform the classical setup with pooling layers \cite{Springenberg2015}.

	Using this concept, one usually builds the encoder by start stacking convolutional layers. Normally, the dimension is already reduced in the first layers by adding a stride or pooling layers. Furthermore, there are dense layers in use when getting towards the latent space. The decoder reverses this process by first enlarging the dimension with dense layers and later on using transposed convolutional layers to get back the dimension of the real space. A transposed convolutional layer is the inverse of a convolutional layer which essentially means that it can decode the features of the input \cite{Zhang2018}.

	\begin{figure}[!ht]
		\centering
		\includegraphics[scale=0.39]{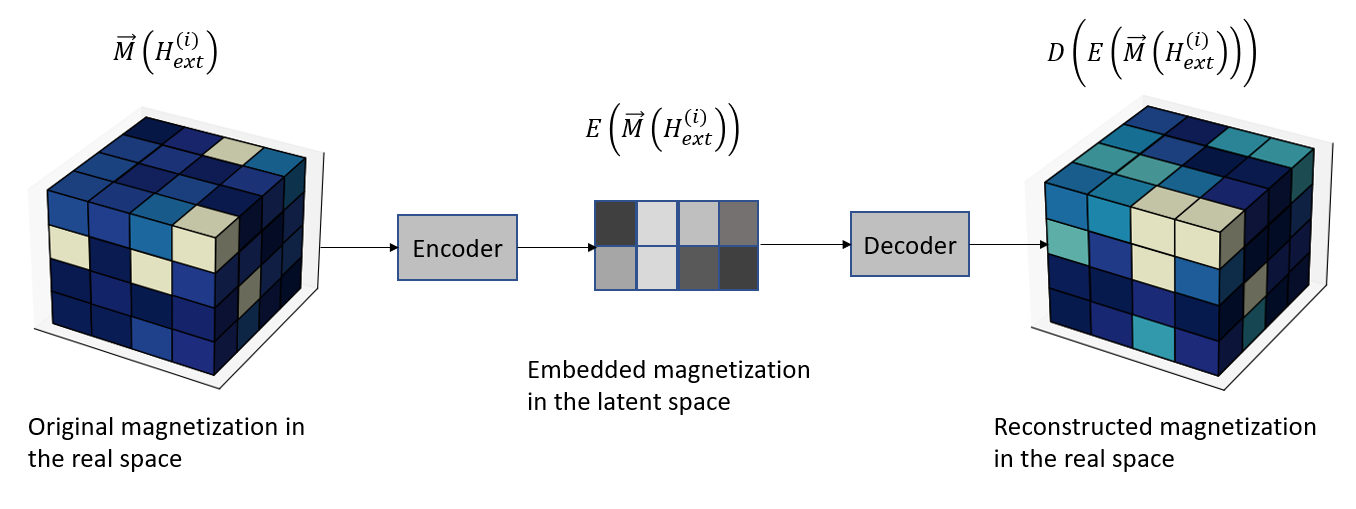}
		\caption{\label{fig:autonecoder} Schematic overview of the reconstruction of one magnetization state by using the autoencoder. Reconstructing a complete hysteresis requires to apply this method to every state along the demagnetization curve.}
	\end{figure}
	
	\section{Results}
	
	\subsection{Data sets}
	
	The dataset used for the following experiments consists of simulated data without practical experiments. The microstructures are generated by Neper which is a tool for generating polycrystalline structures and meshing. In detail, each structure is three dimensional and it consists of 64 cubic grains aligned uniform in all directions. Furthermore, all grains have the same intrinsic magnetic properties besides their easy axes. The easy axes are randomly defined within an angle of 35 degree to the z-coordinate. In addition, the \textit{embedded Stoner-Wohlfarth method} is used to simulate the magnetization reversal process of the microstructures. In detail, the external field is varied from 8 to -8 Tesla in 1000 steps. This means that the external field is reduced by -0.016 Tesla at each field step. The complete dataset consists of 10 full simulation which leads to 10 000 different magnetization states. For each full simulation, a different random distribution of the easy axes is used. From this set nine simulations are used for training while one simulation is the testset. 
	
\begin{figure}[!ht]
		\centering
		\includegraphics[scale=0.31]{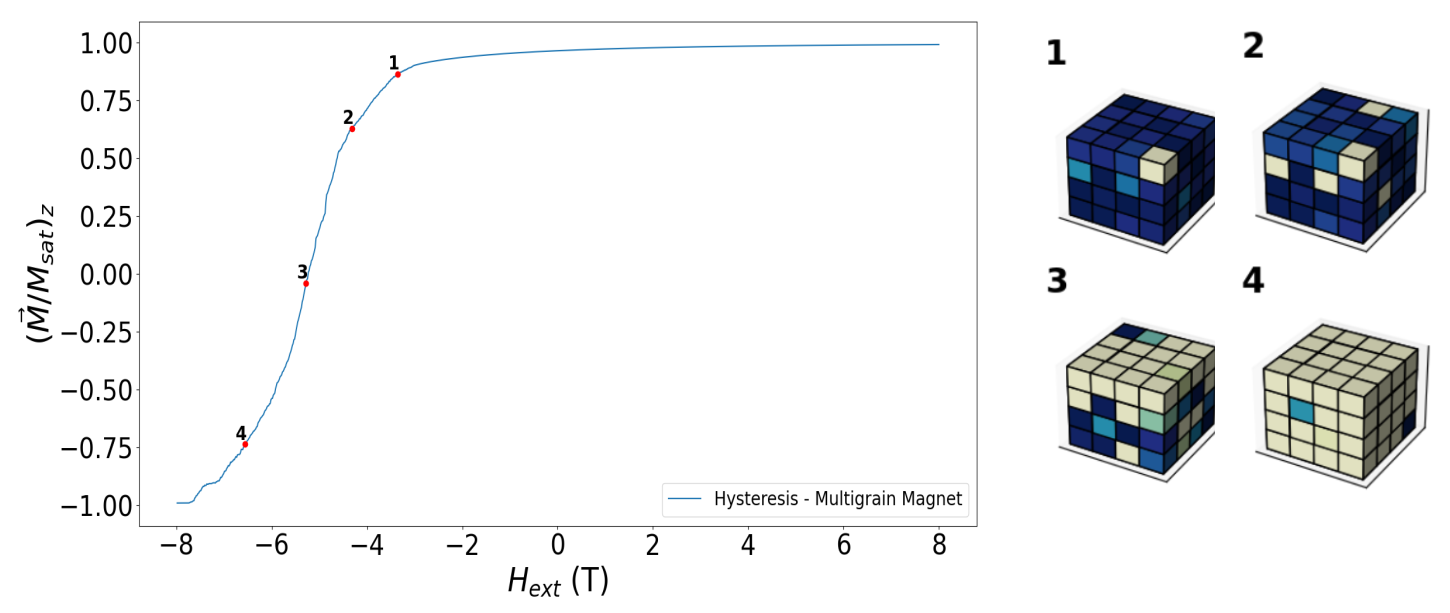}
		\caption{\label{fig:simulatedhysteresis} Simulated magnetization reversal process of multigrain magnet, color shows the magnetization along the z-axis.}
	\end{figure}

	Figure \ref{fig:simulatedhysteresis} shows the simulated magnetization reversal process of one microstructure. Furthermore, there are four magnetization states of the structure visualized explicitly. To recap, this simulation corresponds to 1000 samples of the data set. 
	
	Figure \ref{fig:simulatedhysteresis} indicates that the grains are switching in the range of -8 to -3 Tesla because the magnetization of the complete microstructure does not change significantly before. This essentially means that only approximately 40 \% of the training data contains grains with a reversed magnetization. To mitigate that problem, the training samples are weighted due to the change of the $z$-component of the magnetization state in relation to the previous external field step. The following formula provides the calculation scheme of the weights:
	
	\begin{equation}\label{eq:autoweighttrainingsamples}
	\eqalign{w_1 = 0\textrm{,} \cr
	\eqalign w_i = \sum_{j=1}^{64} |m_{j, z}(H_{ext}^{(i)}) - m_{j, z}(H_{ext}^{(i-1)})| \textrm{ for } i = 2, ... 1000 \textrm{.}}
	\end{equation}

	Here $m_{j, z}(H_{ext}^{(i)})$ denotes the magnetization along the z-axis of the $j$-th grain at the $i$-th external field step. 
	Equation (\ref{eq:autoweighttrainingsamples}) assigns larger weights to training samples which have a significant change in magnetization in relation to the previous field step. This further has the effect that the weights and biases of the models are more sensible to the training samples which are partially reversed. In total this should improve the performance of the models in the part of the demagnetization curve where the magnetization is reversed. 
	
	The implementations of the machine learning models are done with the machine learning framework Tensorflow 2 \cite{Abadi2015} used in Python 3.8. This is an open source software package which was initially developed by Google and it provides a set of software tools to efficiently use machine learning.
	Furthermore, the activation function for all layers is the exponential linear unit \cite{Clevert2016}. In addition, the kernels are initialized by the Tensorflow 2 function 'he\_normal' which draws the initial weights from a truncated normal distribution around 0 where the standard deviation is related to the number input weights. He et. al. have shown that this kernel initialization increases the models accuracy in their work \cite{He2015} in 2015.
	
	The training of the network is done by backpropagation \cite{Schaffer2021} where 20\% of the training samples are used for validation with a batch size of 64. Moreover, the Nadam algorithm is used as the optimizer. The initial learning rate of the method is set to 0.0001. In addition, this parameter of the optimizer is reduced during training by the Keras callback 'ReduceLROnPlateau' if the monitored validation loss stays constant though a certain amount of epochs. Additionally early stopping is used to avoid overfitting. This means that if the validation loss does not decrease significantly after some epochs, the training is stopped before reaching the maximal number of iterations through the training samples.

	\subsection{Encoding magnetic states}
	
	The first question to answer is how to define an appropriate empirical risk minimization problem for training the autoencoder. Since this is a regression problem, the standard mean squared error is a reasonable loss function to choose. In this case, this error is calculated as the sum of squared differences of each entry of the magnetization vector. 
	However, micromagnetism also states that the length of the magnetization vector must be constant. This implies that the loss function can be enriched by a term which takes care of the preservation of the norm. Using these ideas, one can formulate the candidate loss function by the following equation \cite{Kovacs2019}:
	
	\begin{equation}\label{eq:autoencoderlossmsenorm}
	\mathit{L}_{AE}(\hat{x}, x) = a_1\underbrace{\sum_{i,j,k = 1}^4 \sum_{l=1}^3 (x_{ijkl} - \hat{x}_{ijkl})^2}_{\mathit{L}_{MSE}} + a_2 \underbrace{\sum_{i,j,k = 1}^4 \left( 1 - \sqrt {\sum_{l=1}^3  \hat{x}_{ijkl}^2}\right)^2}_{\mathit{L}_{Norm}}   \textrm{,}
	\end{equation}

	where $\hat{x}$ is the reproduced image in the real space of the autoencoder. The indices $i, j$ and $k$ number the grains on a regular geometric grid. Furthermore, $l$ describes the magnetization component. Therefore $x_{ijkl}$ defines the $l$-th component of the unit vector of the magnetization in grain $ijk$. 
	
\begin{figure}[!ht]
		\centering
		\includegraphics[scale=0.36]{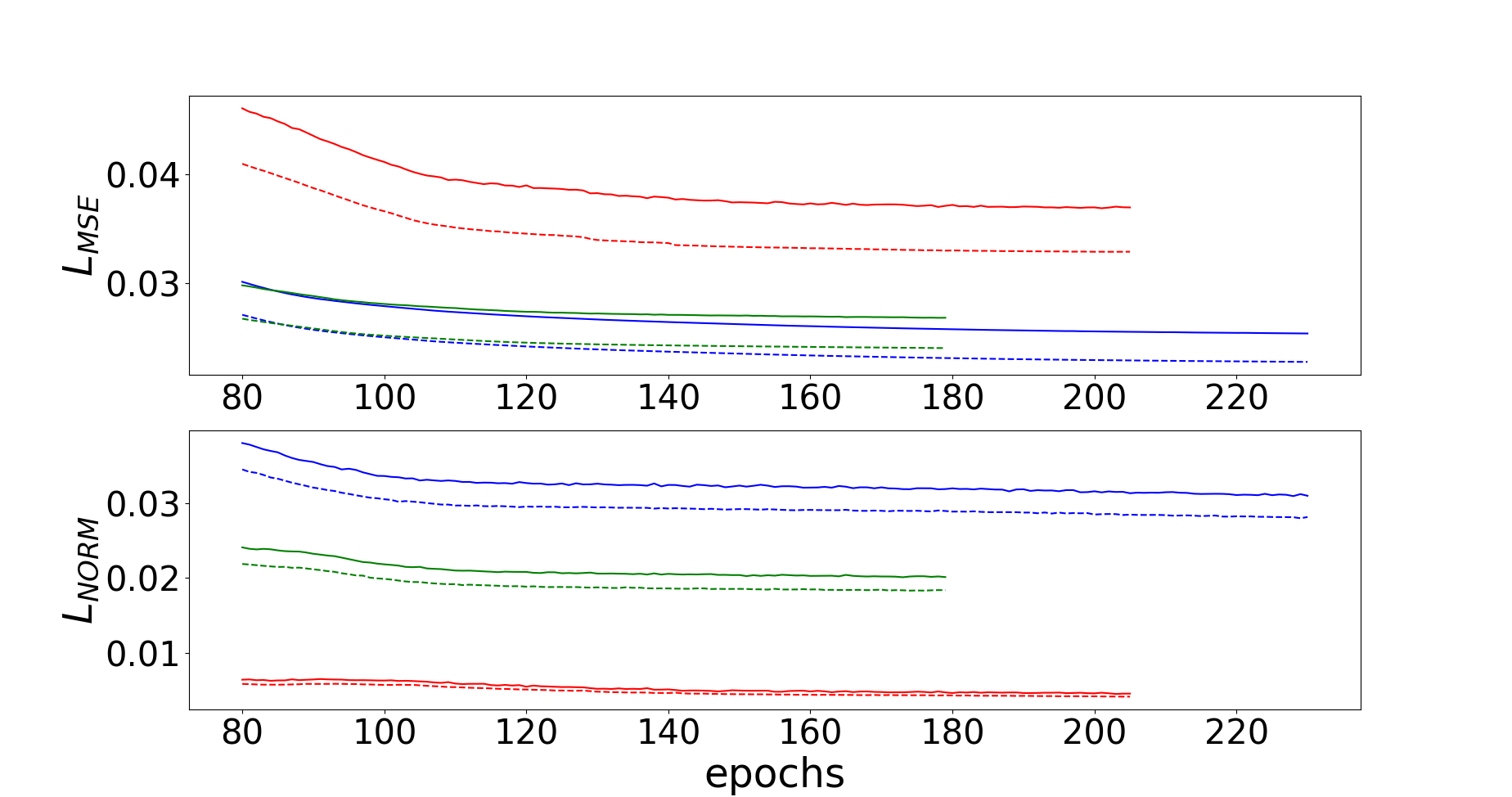}
		\caption{\label{fig:autoencoderloss} Autoencoder loss contributions for different weights $a_1$ and $a_2$. Blue: $a_1 = 1.00$ and $a_2=0.00$; Green: $a_1 = 0.85$ and $a_2=0.15$; Red: $a_1 = 0.50$ and $a_2=0.50$.}
	\end{figure}

	The first term of the right hand side of equation (\ref{eq:autoencoderlossmsenorm}) defines the mean squared error for this example. The second term of this formulation penalizes the model if the reproduced image does not conserve the norm. Furthermore, the parameters $a_1$ and $a_2$ describe the relative weights of the terms since the model eventually produces better results if they are not uniformly weighted. Last, it is worth to note that the loss function accounts only the mean squared error if the parameter $a_2$ is set to zero. Figure \ref{fig:autoencoderloss} compares loss functions for the convolutional autoencoder depending on $a_1$ and $a_2$.
	
	The architecture of the convolutional autoencoder used for the following experiments is described in Table \ref{tab:autoencoderlosstests}. The encoder consists of one convolutional layer with four filters where each filter has the dimension of $2\times2\times2$ and it uses a stride of 2. In addition, there are three dense layers. The latent space consists of 16 neurons which means that the dimensionality is reduced by a factor of 12 since the image is described by 192 floats in the real space. The decoder applies the same setup of layers in the reversed order. In total this leads to 4983 trainable parameters.     
	
	\begin{table}[!ht]
		\centering
		\begin{tabular}{lll} 
			Layer (type)                   & Output Shape                & Param \#    \\ \hline \hline 
			input (InputLayer)           & [(None, 4, 4, 4, 3)]        & 0          \\ \hline 
			conv3d (Conv3D)                & (None, 2, 2, 2, 4)          & 100        \\ \hline 
			flatten (Flatten)              & (None, 32)                  & 0          \\ \hline 
			dense (Dense)                  & (None, 32)                  & 1056       \\ \hline 
			dense\_1 (Dense)                & (None, 32)                  & 1056       \\ \hline 
			dense\_2 (Dense)                & (None, 16)                  & 528        \\ \hline 
			dense\_3 (Dense)                & (None, 16)                  & 272        \\ \hline 
			dense\_4 (Dense)                & (None, 16)                  & 272        \\ \hline 
			dense\_5 (Dense)                & (None, 32)                  & 544        \\ \hline 
			dense\_6 (Dense)                & (None, 32)                  & 1056       \\ \hline 
			reshape (Reshape)              & (None, 2, 2, 2, 4)          & 0          \\ \hline 
			output (Conv3DTransp) & (None, 4, 4, 4, 3)          & 99         \\ \hline \hline 
			Total params: 4,983 \\ 
			Trainable params: 4,983 \\ 
			Non-trainable params: 0 \\ \hline 
		\end{tabular} 
		\caption{\label{tab:autoencoderlosstests} Autoencoder architecture for evaluating different loss functions.}
	\end{table}
	
	Figure \ref{fig:autoencoderloss} shows the mean squared error and norm error dependent on the number of epochs for the different loss functions. Note that the terms are separately tracked without weighting during the training which ensures that they are quantitatively comparable for the different setups.
	The graphs indicate that the model does not loose the accuracy in terms of the mean squared error by adding the penalization term. This means that including the norm in loss function reduces the mean squared error simultaneously. Nevertheless, it is necessary to mention that this behavior is only shown if the norm related term is weighted smaller than the term which calculates the mean squared error. In numbers this means that setting the relative weights $a_1$ to 0.85 and $a_2$ to 0.15 enables the model to reach approximately the same minimum of the mean squared error by less epochs. However, it is worth to note that weighting the norm error equally to the mean squared error introduces a significantly higher mean squared error. 
	
	As expected, adding the physical information of the preservation of the norm to the loss function significantly reduces the error in the length of the predicted magnetization vectors. In addition, the relative difference between the norm error on the training set and the norm error on the validation set decreases by enlarging the relative weight $a_2$. However, the preservation of the norm is just additive information since this term does not relate the predicted magnetization to the ground truth. This means that weighting the norm error to high would lead bad predictions since the model would not learn the underlying data distribution in any sense.   
	
\begin{figure}[!ht]
		\centering
		\includegraphics[scale=0.36]{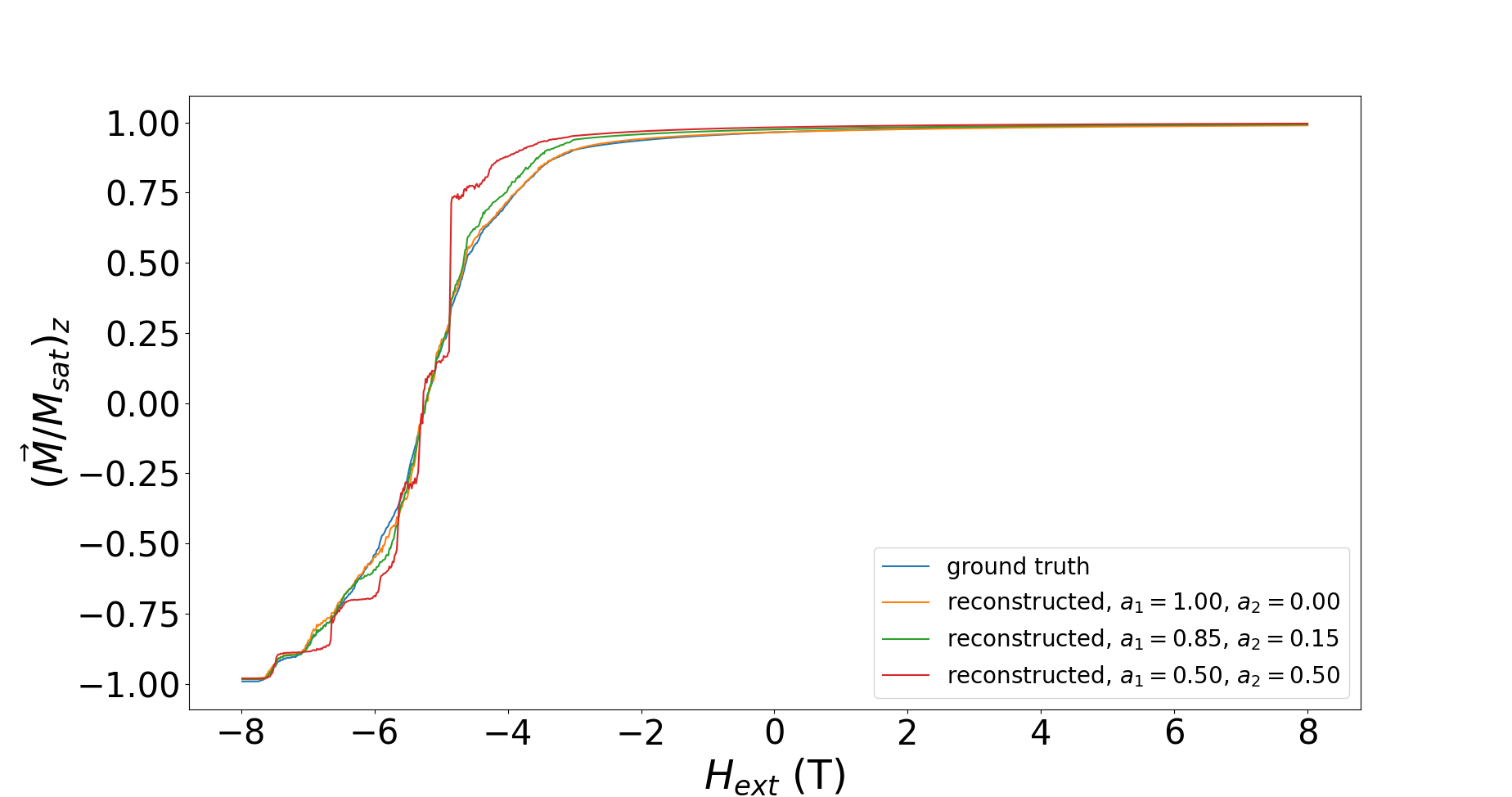}
		\caption{\label{fig:autoencoderdemagcurve} Ground truth and reproduced magnetization states of the convolutional autoencoder for different weights $a_1$ and $a_2$. Orange: ground truth; Blue: $a_1 = 1.00$ and $a_2=0.00$; Green: $a_1 = 0.85$ and $a_2=0.15$; Red: $a_1 = 0.50$ and $a_2=0.50$.}
	\end{figure}

	Figure \ref{fig:autoencoderdemagcurve} shows the magnetization reversal process of the test microstructure simulated by the \textit{embedded Stoner-Wohlfarth method}. In addition, there are the plots of the reproduced magnetization of the structure where the autoencoder is trained with the different loss functions which are analyzed before. The graphs are created by predicting the magnetization state of all steps of the external field and plotting the $z$-component correspondingly.  
	
	In general the graphs indicate that the used model is able to reproduce the demagnetization curve. Furthermore, one can see that the mean squared error is a good error measurement for the underlying problem. Nevertheless, there are differences in the performance related to the configuration of the loss function. In detail, the plots show that using a model that minimizes a loss which equally weights the norm error and the mean squared error performs worse than a model which weights the mean squared error higher than the norm error. However, weighting the norm conservation term by 15\% provides a model with a reasonably good reconstruction.
	Nonetheless it is worth to note that the model in this test case gets inputs with a normalized magnetization. So for this case, the autoencoder predicts accurately if it does not change the norm but this is eventually not true if there is an additional machine learning model which predicts the magnetization state of the microstructure in the latent space. If the predictor produces embedded images which are not normalized in the real space, the autoencoder must be able to do this normalization. In total this is an additional argument to add the norm error to the loss function since the mean squared error on its own does not include normalization. 
	
 	\subsection{Predictor}
 	
 	The autoencoder can be used for dimensionality reduction but the main goal is to predict the complete magnetization reversal process of unseen microstructures. This means that the regressor must be able to predict accurate results recursively from its own predictions. 
 	In general, the dataset to train and test the latent space predictor is the same as for the autoencoders but it needs to be preprocessed. In detail, the first thing to do is to choose a single model for the dimensionality reduction. In the following experiments this is the autoencoder given by Table \ref{tab:autoencoderlosstests} with the loss weights $a_1 = 0.85$ and $a_2 = 0.15$. The encoder of this model is used to obtain the embedded images of all states of magnetization of the 10 simulations. Again one full simulation is taken as the unseen test set to evaluate the models. 
 	The next step is to concatenate the training samples in batches of $t$ consecutive magnetization states. To avoid misunderstandings, it is worth to note that one magnetization state can be part of at most $t$ different input batches. In addition, the strength of the external field at the $(i-1)$-th step is added to these batches. In total this leads to $(1000-t)$ inputs for each simulation. 
 	
 	\begin{figure}[!ht]
 		\centering
 		\includegraphics[scale=0.25]{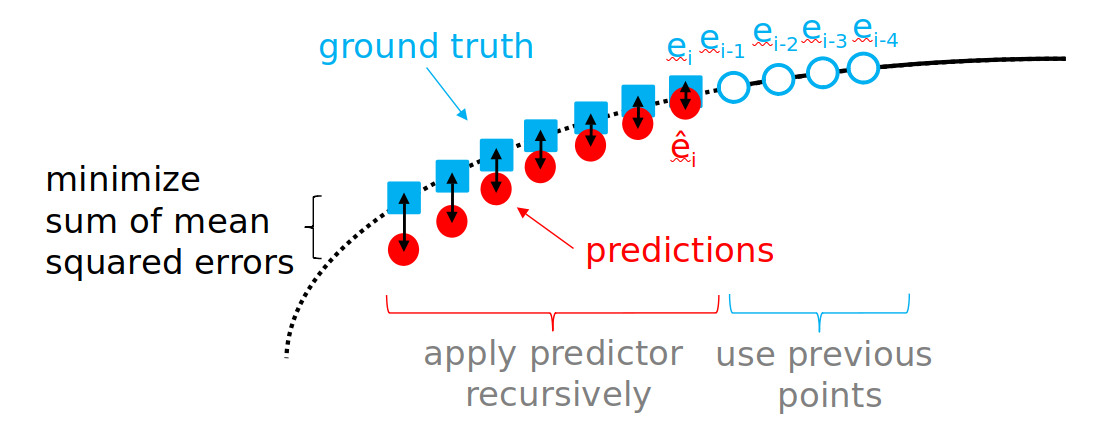}
 		\caption{\label{fig:predtrainingsketch} Training process of a predictor model with t=4 and l=7.}
 	\end{figure}
 	
 	However, the next question to answer is how to define a loss function for an empirical risk minimization problem for the model. In general the predictor should be able to estimate the complete demagnetization curve and not only the next magnetization state. Therefore it must be able to provide accurate results even if the inputs are magnetization states which are calculated by the model itself. This is the reason why a standard mean squared error loss function which only measures the difference to the next state in magnetization eventually fails due to interfering errors. To avoid this problem, one can use a loss function which also accounts for a mean squared error of further predictions. In summary one training sample of the predictor has the following structure: 
 	
 	\[b^{(i)} := (H_{ext}^{(i-1)}, e^{(i-t)}, ... e^{(i)}, e^{(i+1)}, ..., e^{(i+l-1)}) = 	\] 
 	\[ =(H_{ext}^{(i-1)}, E(\vec{M}(H_{ext}^{(i-t)})), ..., E(\vec{M}(H_{ext}^{(i)})), E(\vec{M}(H_{ext}^{(i+1)})), ..., E(\vec{M}(H_{ext}^{(i+l-1)}))) \textrm{,} \]
 	
 	where $e^{(i)}:=E(\vec{M}(H_{ext}^{(i)}))$ denotes the encoded magnetization of the $i$-th external field step. The $i$-th batch contains the encoded magnetization patterns of the $t$ previous field steps and the $l$ future field steps. 
 	Furthermore, the following notation is introduced: 
 	
 	\begin{equation}\label{eq:predehatnotation}
 	\epsilon^{(j)} :=
 	\cases{ e^{(j)} \quad \textrm{ if } j<i \\ \hat{e}^{(j)} \quad \textrm{ else } }{\textrm{.}}
 	\end{equation}
 
 	Here $\hat{e}^{(j)} := P(H_{ext}^{(i)}, \epsilon^{(j-1)}, ..., \epsilon^{(j-t)})$ is the output of the predictor at the j-th step of the demagnetization curve. Using these notations, the mean squared error of one prediction is given by the following formula: 
 	 	
 	\begin{equation}\label{eq:predictoronepredloss}
 	\mathit{L}_{j}(\hat{e}^{(j)}, e^{(j)}) = \sum_{t = 1}^d (e^{(j)}_t - \hat{e}^{(j)}_t )^2 \textrm{,}
 	\end{equation}
 
 	where $d$ is the number of latent space variables. Finally, it is necessary to sum up the mean squared errors for each prediction to get the loss function: 
 	\begin{equation}\label{eq:predictorloss}
 	\mathit{L}_{PRED}(\hat{e}, e) = \sum_{j = 0}^{l-1} \mathit{L}_{j}(\hat{e}^{(j)}, e^{(j)}) \textrm{.}
 	\end{equation}
 	
 	Figure \ref{fig:predtrainingsketch} outlines the training process of the predictor graphically. The sketch uses a model where $t=4$ and $l=7$. In this case, the neural network takes the four previous magnetization states as an input. Furthermore, it is applied recursively to estimate the next seven magnetization states and the sum of the mean squared error of these predictions is minimized during the training process.

 	\begin{figure}[!ht]
 		\centering
 		\includegraphics[scale=0.36]{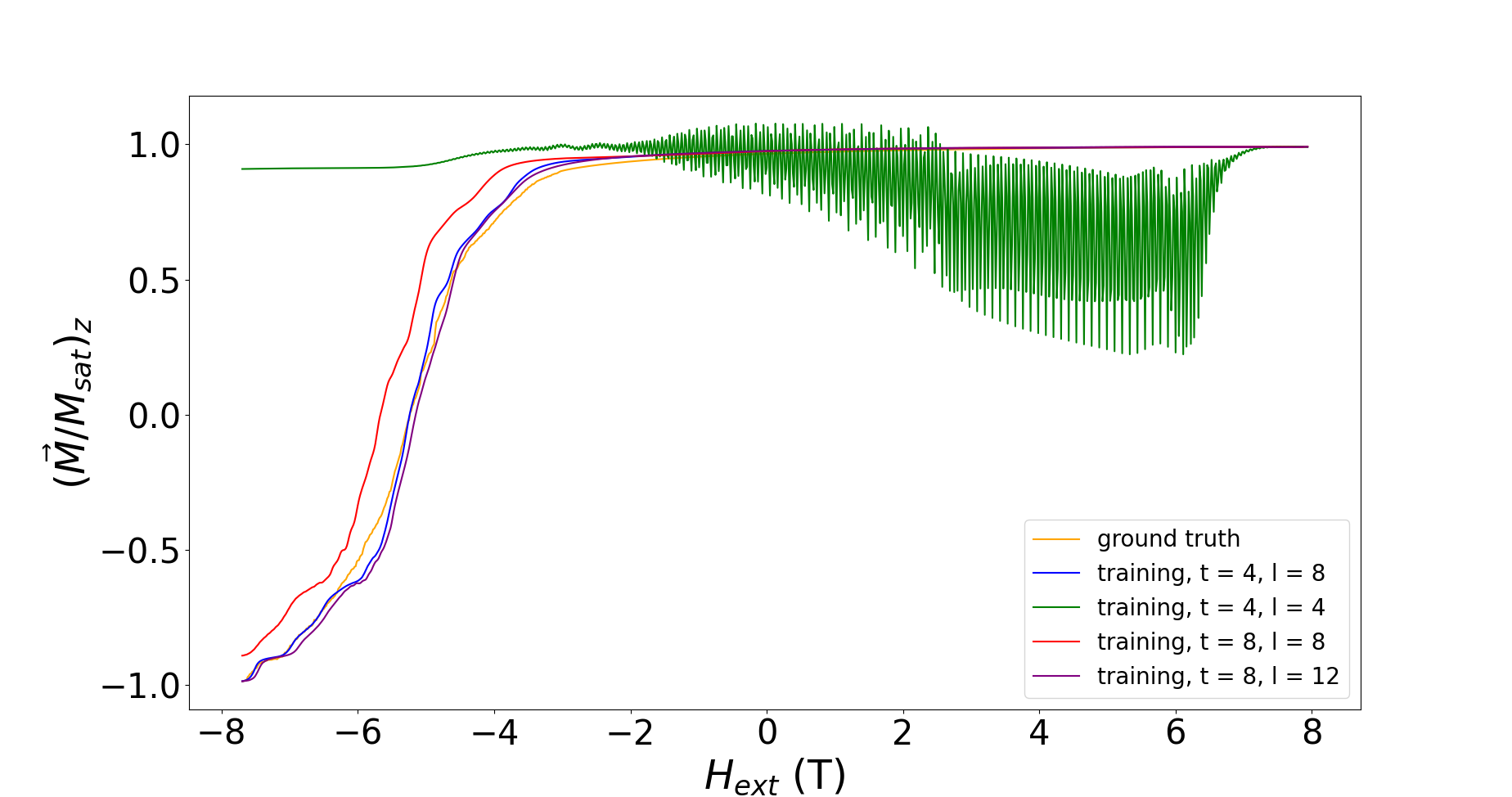}
 		\caption{\label{fig:predtlhysteresis} Demagnetization curve of microstructure calculated by the \textit{embedded Stoner-Wohlfarth method} and predicted by latent space regressor with different values of $t$ and $l$.}
 	\end{figure}

 	Figure \ref{fig:predtlhysteresis} shows the complete demagnetization curve of the test microstructure predicted by the model with different configurations of $t$ and $l$. In addition, there is the ground truth which is the calculated hysteresis by the \textit{embedded Stoner-Wohlfarth method}. The predictor model takes as an input the first $t$ encoded magnetization states from the test set and calculates the complete magnetization reversal recursively.
 	
 	Although the minimization of the loss function ensures that each of the configurations is able to do at least 4 predictions with sufficiently small error, it is clearly displayed that $l$ must be greater than 4 to enable the model to calculate the complete demagnetization curve reasonably well. In detail, the model where $t=4$ and $l=4$ does not reproduce the hysteresis and the oscillating predictions are eventually related interfering errors. However, if $l=8$ or greater the model is able to estimate the curve of interest. Nevertheless, it is interesting to see that the configuration $t=8$ and $l=8$ scores worse than the model with $t=4$ and $l=8$. In total the simulations show that the configurations where $t$ is is at least one third less than $l$ outperform the other combinations. 
 	
	\section{Summary and conclusion}
 		
	This work outlines a machine learning approach to estimate the demagnetization reversal process of multigrain microstructures. The method uses a convolutional autoencoder and a latent space regressor based on neural networks. Furthermore, the \textit{embedded Stoner-Wohlfarth method} is used to generate the dataset. This numerical method is efficient enough to calculate a dataset to empower deep learning for predicting the hysteresis loop of an unseen microstructure. 
	
	The analysis shows that it is possible to reproduce the demagnetization curve with an autoencoder trained by the loss function given by equation (\ref{eq:autoencoderlossmsenorm}). However, the dominating term of the loss function must be the mean squared error since weighting the norm related term too high leads to worse reconstructions. Nevertheless, the results indicate that including the norm error reduces the computational expense for training the model. Furthermore, it eventually includes normalization capabilities to the autoencoder. 
	
	In summary the experiments with the latent space predictor prove that it is possible to predict the hysteresis curve of an unseen microstructure by a machine learning model operating in the latent space of an autoencoder. However, there are some important facts which need to be fulfilled to enable the method to succeed. The first is that the model based on neural networks must have enough hidden layers to estimate the underlying distribution. Furthermore, it is necessary to recursively apply the predictor during training by using a loss function which takes the mean squared error of future predictions into account. Moreover, simply increasing the number of inputs of previous magnetization states does not increase the model performance as long as the number of future predictions within the loss function is not changed accordingly. The discussed results of the predictor take as an input the first $t$ encoded magnetization states. However, different starting points lead to the same results which most likely is related to the fact that the training samples are differently weighted. To recap, datapoints where irreversible magnetization reversal occurs have a higher impact to the machine learning models than datapoints where only reversible magnetization rotation occurs.

	A key finding of the work is that loss function engineering is necessary to apply common machine learning techniques to micromagnetism. One insight which was gained during this work is that adding the norm error to the loss function of the autoencoder did provide better results for this setup than just taking the standard mean squared error. In detail, adding this term with an appropriate weight led to the same results for the mean squared error while reducing the error in the norm. Moreover, the inclusion of this term reduced the training time of the model since it was able to find a minimum of the loss functions after less epochs. Furthermore, we were able to find differences between the ground truth and the reproduced images which gives room for improvement for this method.
	
	In addition the experiments with the predictors clearly stated the fact that it is necessary to account the error of future prediction during the training of the model to enable it to predict the complete hysteresis curve. Therefore a recursive loss function was developed which respects the mean squared error of the future predictions. Furthermore, we were able to show that the neural network for calculating the following steps of the demagnetization curve must be deep enough to approximate the magnetization states accurately. Finally, we evaluated architectures for the predictor by trying different combinations of the hyperparameters which describe the number of input magnetization states and the number of recursive iterations during the training. These experiments showed that ratio between these parameters effects the accuracy of the model. 
	
	The goal of this work was a proof of concept of the machine learning method for calculating the demagnetization curve of magnets with simple granular geometry. Nevertheless, it is possible to extend this method for synthetic polycrystalline material. However, more complex structures eventually need more complex methods. In detail, the experiments with the norm error showed that adding information to the autoencoder improves the model. Therefore it would be interesting to apply additional information to the loss function. This can be a term which penalizes the model if the predicted state is not in its equilibrium. This idea comes from the work of \cite{Kovacs2022}  which introduces physical informed neural networks where the underlying equations of micromagnetism are part of the loss function. Moreover, another promising way to improve the predictor is to enrich its input by information about the microstructure and the intrinsic magnetic properties of the material. However, since high dimensionality is expected to be a problem one has to find a way to represent the microstructure itself within the latent space.

	\section*{Acknowledgments}
	Financial support by the Austrian Federal Ministry for Digital and Economic Affairs, the National Foundation for Research, Technology and Development and the Christian Doppler Research Association is gratefully acknowledged, as well as the support by the Austrian Science Fund (FWF) under grant No. P31140-N32 and grant No. F65.
	\newline\\

	\bibliography{bibref}
	
\end{document}